\documentclass[11pt]{article}
\usepackage{amsmath}
\usepackage{latexsym}
\usepackage{amssymb}

\numberwithin{equation}{section}
\title{Yang-Mills Theory In, Beyond, and Behind Observed Reality
\footnote{Solicited contribution to the volume ``Fifty Years of Yang-Mills Theory'' 
(WorldScientific).}}
\author{Frank
Wilczek\footnote{wilczek@mit.edu}}
\begin{document}
\maketitle

\begin{abstract}
The primary interactions of Yang-Mills theory \cite{ym} are visibly embodied in hard processes, 
most directly in jets.   The character of jets also reflects the deep structure of effective charge, 
which is dominated by the influence of intrinsically nonabelian gauge dynamics.   
These proven insights into fundamental physics ramify in many directions, 
and are far from being exhausted.    
I will discuss three rewarding explorations from my own experience,
whose point of departure is the hard Yang-Mills interaction, and whose end is not yet in sight.   
Given an insight so profound and fruitful as Yang and Mills brought us, 
it is in order to try to consider its broadest implications, which I attempt at the end.  

\end{abstract}

\section{Yang-Mills Theory, Directly in the Phenomena}

The historical path whereby a particular Yang-Mills theory, 
quantum chromodynamics or QCD, was established as the theory of the strong interaction 
\cite{dgross1973, polit1973, dgrossb1973, dgross1974, hgeorgi1974}
was rather complex.  It came through the analysis of deep inelastic scattering, 
and involved both intricate kinematics (dispersion relations for moments of structure functions) 
and elaborate theoretical machinery (operator product expansion, renormalization group for 
Wilson coefficients).   

Things might have developed quite differently if higher-energy accelerators, 
specifically the CERN Large Electron-Positron Collider (LEP) had come into operation earlier.  
{}For in the phenomena observed at LEP the crucial aspects of QCD -- 
its limbs, the joints, and soul -- or, more prosaically, its basic constituents, its primary interactions, 
and its central new dynamical feature, asymptotic freedom -- 
all become, to the awakened mind, easily visible realities.  

Indeed, there are two broad classes of events observed at LEP, each occurring roughly half the time.  
The first class involves leptons and photons.  Within this class, about 99\% of the events consist simply of 
two tracks -- a charged lepton and its antiparticle emerging in opposite directions.    
In the remaining 1\%, those particles are accompanied by a photon that carries off a substantial 
fraction of the energy and momentum.  (There are also, very rarely, multiple photon emissions.)  
These events, of course, reflect the structure of quantum electrodynamics (QED).  
Photon radiation is a direct manifestation of the basic interaction vertex.  
The entire theory of QED can be built up from this interaction vertex, 
strictly by following the algorithms of quantum field theory -- 
i.e., the logical implementation of special relativity and quantum mechanics.   
By measuring the angular and energetic distributions of the photons  
-- their antenna pattern -- we can examine the responsible interaction minutely, 
and thereby monitor, and check the soundness of, QED's heartbeat.  

The second class involves strongly interacting particles.  
Within this class, about 90\% of the events consist of two ``jets'' moving in opposite directions; 
about 9\% contain three jets; .9\% four jets, and so on \cite{aleph, opal, L3}.  
The jets here are not single particles, 
like leptons or photons.  Rather they are sets of several strongly interacting particles 
-- hadrons like protons, pions, and their many, often highly unstable brethren and their antiparticles --  
moving rapidly in the nearly the same direction.  

Upon looking at the patterns of tracks recording events of the 
second class, you can't help but be strkiking their striking qualitative resemblance to the first class.   
Once such images had become available, theorists could hardly have avoided suggesting that 
the jets should be analyzed as effective particles, and that one should try to build up a description 
of the observations using a quantum field theory of these particles, on the model of QED.  

But which theory?  Well, at first people might have tried QED itself, or perhaps some sort of Yukawa theory, 
with scalar particles.    Analyzing the three-jet antenna patterns, theyÕd quickly find that the QED description 
worked well.  But beyond that, for four jets or more, it would not.    
Here, then, would come the critical juncture.   If, in this imaginary history, 
the images had arrived before 1954, people would have had to come up with an essentially new theory.  
But by 1954 the right framework was in hand.  It is, of course, Yang-Mills theory \cite{ym}.   
By detailed quantitative work, physicists would have arrived at the gauge group $SU(3)$ and the 
triplet assignments of fermions (with fractional electric charge)  -- that is to say, at modern QCD, 
directly from the phenomena.   

In real history the interpretation of jets at LEP served rather as a verification of 
earlier, much more arduous and indirect inferences, rather than a process of discovery.    
But the results are no less beautiful.  To me, indeed, their beauty seems enhanced.  They embody 
profound realization of a brilliant human creation rather than accidental revelation, blindly stumbled upon.  

The interpretation of jets as representing the energy-momentum flow of basic particles, 
of course, begs the question why they are \emph{not} in fact single particles.    
Here is where asymptotic freedom enters the picture.   
One observes that ``hard'' radiation, which substantially changes the overall flow of energy and momentum, is rare; 
but that ``soft'' radiation, which creates new particles while not disrupting the flow of energy momentum, is common.   
The rarity of hard radiation explains the decreasing probability of multi-jet events; 
while the commonness of soft radiation explains why each jet consists of many particles.    
This effect can be quantified -- in our imaginary history, undoubtedly it would have been -- 
and the running of the effective coupling thereby established, directly from the phenomena.  

The weakening of effective couplings at large energy-momentum corresponds, in real space, 
to the weakening of effective charge at short distances.   It means that virtual particles \emph{antiscreen} 
a test color charge.  That is counterintuitive and very unusual \cite{zee, coleman} behavior.    
But it is absolutely crucial to use of QCD to describe the strong interaction.  We need 
to explain how a basic coupling, that is observed to be small at short distances, builds up 
the most powerful force of Nature, which permanently confines those who dare disturb its equanimity!  

Running of the couplings and anti-screening in Yang-Mills theory was first established 
by straight unguided calculation.  Before long, however, it was understood in terms 
more-or-less accessible to intuition  \cite{nielsen1981}.    
I will not rehearse the details of that understanding here; 
it suffices to recall that the color paramagnetism of gluons, associated with their spin, is the key.   
This represents a triumph for the fundamental ideas of Yang and Mills, on two counts: 
the fact that the gluons carry color charge directly reflects the 
nonabelian nature of color $SU(3)$ gauge symmetry; 
and the crucial gyromagnetic ratio $g=2$ is a direct consequence of the minimal Yang-Mills equations.

\section{From Running Coupling to Quantitative Unification to
\\ Supersymmetry}

The calculation of running, of course, extends immediately to
electroweak interactions. (Indeed, my own interest in it largely
originated from this angle.) It was put to brilliant use in the famous
work of Georgi, Quinn, and Weinberg \cite{hgeorgib74}, who indicated 
through its use how
dreams about unification of interactions
\cite{pati1973, fw10} could be brought down to earth.  One could check concretely
whether the observed, unequal couplings might result from running a
single coupling from ultra-short to accessible distances. A few years
later Dimopoulos, Raby, and I  realized \cite{dim1981} -- initially to my great surprise
-- that including the effects of low-energy supersymmetry,
which is quite a drastic expansion of the physics, makes only
comparatively small changes in the predictions that emerge from this
sort of calculation.   Precision experiments and improved
calculations appear to endorse these dreams and ideas, in their
supersymmetric version.

Unless this is a cruel tease on the part of Mother Nature, it means
we can look forward to a lot of fun exploring supersymmetry, and
maybe some aspects of unification, at the LHC. An especially poetic
possibility is to explore the possibility that other sorts of
parameters, besides gauge couplings, derive by running from a unified
value \cite{fw12}. It is widely speculated that the masses of 
different sorts of
gauginos, or of squarks and sleptons, might be related in this
way.

\section{From Dark Momentum to Gluonization to Higgs and Dark Matter}

Feynman interpreted the famous SLAC experiments on deep inelastic
scattering using an intuitive model of nucleons that postulated
point-like particles (partons) as nucleon constituents and treated
their dynamics in a crude impulse approximation, ignoring both
interactions and quantum interference \cite{feyn1970}. Identifying 
the partons as
quarks, and building the weak and electromagnetic currents by minimal
coupling to quarks, led to many successful predictions 
\cite{bjo1969}. There was,
however, one clear failure. The momentum carried by quarks inside a
fast-moving proton does not add up to the total momentum of the
proton, in fact it is less than half.

Today's ``dark matter" problem in astronomy is reminiscent of
that old ``dark momentum" problem. In the formal treatment of deep 
inelastic scattering, the analogy becomes eerily precise. In that
framework, the (failed) sum rule expresses saturation of the full
energy-momentum tensor by the energy-momentum tensor constructed
from quarks \cite{dgross1974, hgeorgi1974}.  But where electroweak 
currents see only quarks, gravitons
see more!  We realized early on \cite{dgross1974, hgeorgi1974} that 
the color gluons of QCD, which
are electroweak singlets but do carry energy-momentum, would enable
us to keep the good predictions while losing the bad one.  Evidently
the gluons had to be major, though ``dark" (or better: invisible),
constituents of the proton.

Our analysis of deep inelastic scattering, which followed pioneering
ideas of Wilson \cite{wil1971}, and built on the insightful hard work 
of Christ,
Hasslacher, and Mueller \cite{christ1972}, went beyond the parton 
model in other, more
profound ways. A fast-moving quark is revealed, to probes at higher
resolution (higher $Q^2$), to be composed of slower-moving (smaller
$x$) quarks, anti-quarks and gluons, which in turn will resolve into
more, softer stuff. This process, seen experimentally as evolution of
structure functions, is deeply characteristic of quantum field theory.

These evolution effects further enhance the role of glue in the
proton.  Several of us worked out that there should be a dramatic
pile-up of soft stuff, particularly soft glue, at small $x$ 
\cite{rujula1974}. To a
hard current (indirectly), or to a hard graviton (theoretically), the
proton mostly looks like a blob of soft glue. After a long interval,
beautiful work at HERA has confirmed these predictions in impressive
detail \cite{wolf}.

Very soft or ``wee" constituents of protons played a major role in
Feynman's ideas about diffractive scattering \cite{feyn1969}. His 
idea was that in
diffractive scattering, by exchange of wee partons, the relative
phases between different multiparton configurations in the proton
wave function get disrupted, without much transfer of
energy-momentum.   These ideas are intuitively appealing, and have
inspired some successful phenomenology, but as far as I know they
haven't yet been firmly rooted in QCD.

Much better understood -- I hope! -- is the importance of
gluonization for some frontier topics in high-energy physics, namely
Higgs particle production and WIMP searches.  The primary, classical
coupling of Higgs particles is to quarks, proportional to their mass.
But because the $u$ and $d$ quarks we mainly find inside nucleons are
so light, their direct coupling is heavily suppressed. Instead the
most important coupling arises indirectly, as a quantum effect,
through virtual top quark loops connecting to two gluons \cite{wilczek1977}.

Originally I was interested in this Higgs-gluon vertex for its
potential to induce Higgs particle decays. Georgi, Glashow, Machacek, 
and Nanopoulos \cite{georgi1978} quickly
realized it could be exploited for production of Higgs particles at
hadron colliders, through gluon fusion. This process, which of
course relies completely on the glue content of protons, is expected
to be the main production mechanism for Higgs particles at the LHC.
It is important to calculate the production rate accurately,
including good estimates of the gluon distribution functions, so that
we will be able to interpret the observed production rate, and check
whether the basic vertex is in fact what the standard model, in this
intricate way, predicts.

The on-shell Higgs particle, due to its large mass, couples to hard gluons.  
When considering detection
of the sorts of dark-matter candidates provided by models of
low-energy supersymmetry, however, we find ourselves in quite a
different kinematic domain. Since these WIMPs will be heavy, and
very slowly moving by particle physics standards, they will scatter at
very small momentum transfer. The coupling of SUSY WIMPs to matter depends
on poorly constrained details of the models, but in many realizations
it is dominated by virtual Higgs exchange. Here the Higgs-gluon
vertex comes in at essentially zero energy-momentum. Shifman,
Vainshtein and Zakharov \cite{shifman1978}, in beautiful work, 
related the relevant
gluon operator to the trace of the energy-momentum tensor, whose
matrix elements are of course known. Their reasoning brings us right back to the old dark
momentum problem, thus closing a full circle.

It is philosophically profound, and quite characteristic of modern
physics, that even when viewing something so basic and tangible as a
proton, what you see very much depends on how you choose to look.
Low-energy electrons see point-like particles, the version described
in old high-school textbooks; hard currents see an evolving pattern
of quarks; gravitons see these plus lots of gluons as well; wee
gluons see some complicated stuff we don't properly understand (we do
know its name, Pomeron); real Higgs particles see gluons almost
exclusively; and WIMPS, through exchange of virtual Higgs particles,
see the Origin of Mass!  {}For the trace of the energy momentum tensor, to
which they mainly couple, is on the one hand dominated by contributions from massless
color gluons and nearly massless quarks, and on the other hand equal to the nucleon mass. 
Each probe
reveals different aspects of a versatile reality.

\section{From Asymptotic
Simplicity to Quark-Gluon Plasma to Quark-Hadron Continuity}

Over the years we've learned
to use the concept of asymptotic freedom more boldly and confidently.  
To put it differently, we've learned fruitful ways
to lower our standards.   Instead of trying to prove directly from first principles
that weak coupling applies, we usually
content ourselves with consistency checks. That is, we tentatively
assume that weak coupling calculation of some quantity of interest
starting with quark and gluon degrees of freedom is adequate, and
check whether the calculation contains infrared divergences 
\cite{sterman1977}. This
check is by no means trivial, since QCD is full of massless (color)
charged particles. So in cases where we find there are no infrared
divergences we declare a well-earned victory, and anticipate that our
calculations will approximate reality.  This strategic retreat has
licensed a host of successful applications to describe jet
processes, inclusive production, fragmentation, heavy quark physics,
and more.

We aren't always forced to compromise. In some important
applications, including low-energy spectroscopy, direct integration
of the equations using the techniques of lattice gauge theory is
practical. But as physicists hungry for answers, we properly regard
strict mathematical rigor as a desirable luxury, not an indispensable
necessity.

A particularly interesting and important application of the looser 
philosophy is to construct self-consistent descriptions of extreme 
states of matter, starting from quarks and gluons \cite{QCD2001}.

The high temperature, low baryon number regime is foundational for
very early universe cosmology.  It is also the object of an intense,
international experimental program in relativistic heavy ion physics.
The overarching theme is that a weak coupling description of
high-temperature matter, starting with free quarks and gluons,
becomes increasingly accurate as the temperature increases. This can
be seen, for the equation of state, from numerical simulation of the
full theory \cite{skatz}. After heroic calculations, which introduce several
ingenious new techniques, controlled quasi-perturbative calculations
(including terms up to sixth order in the coupling, and some infinite
resummations) match the numerical work \cite{yschroder}. This is a milestone
achievement in itself, and also promising for future developments,
since the weak coupling techniques are more flexible. They might be
applied, for example, to calculate viscosity and energy loss, which
can be probed experimentally. In this way, we can hope to do justice
to the vision of quark-gluon plasma.

The regime of high baryon number density, and low temperature, is
intrinsically fascinating, and might be important for describing the
inner dynamics of supernovae and the deep interior of neutron stars.
The first fundamental result about QCD at high baryon number density
is that many of its key properties, including for example the
symmetry of the ground state and the energy and charge of the
elementary excitations, \emph{cannot} be calculated to a good
approximation starting from fermi balls of non-interacting quarks.
The perturbation theory (for just about anything) contains infrared
divergences \cite{QCD2001}.

Fortunately, the main source of these divergences is well understood.
They signal instability toward the development of a condensate of
quark pairs, similar to the Cooper pairs that occur in metallic
superconductors. Whereas the phenomenon of superconductivity in
metals is very delicate, because one must overcome the dominant
Coulomb repulsion of like charges, color superconductivity is very
robust, because there is a fundamentally attractive force between
quarks (in the color and flavor antitriplet, spin singlet channel).
One can construct an approximate ground state that accommodates the
pairs, adapting the methods of BCS theory. Perturbation theory around
this new ground state no longer has infrared divergences. Thus we
find that strongly interacting matter at asymptotically high density
can be studied using weak coupling, but non-perturbative methods.

Color superconductivity has become an extremely active area of
research over the past few years, and many surprises have emerged.
Perhaps the most striking and beautiful result is the occurrence of
color-flavor locking, a new form of symmetry breaking, in real-world
(3 flavor) QCD at asymptotic densities \cite{malford1999}. The symmetry
$SU(3)_C \times SU(3)_L \times SU(3)_R$ of local color times chiral flavor is broken
down to the diagonal subgroup, a residual global $SU(3)$.

Color-flavor locking is a rigorous, calculable consequence of QCD at
high density. It implies confinement and chiral symmetry breaking.
The low-energy excitations are those created by the quark fields,
those created by the gluon fields, and the collective modes
associated with chiral symmetry breaking. Because CFL ordering mixes
up color and flavor, the quarks form a spin-1/2 octet (plus heavier
singlet), the gluons form a vector octet, and the collective modes
form a pseudoscalar octet under the residual $SU(3)$.
Altogether there is an uncanny resemblance between the properties of
dense hadronic matter one calculates for the CFL phase, and the
properties one might anticipate for ``nuclear matter" in a world with
three massless quarks.

A nice perspective on this arises if we consider elecromagnetic properties, which we can do 
by coupling in the $U(1)$
of electromagnetism. Both the original color gauge symmetry and the
original electromagnetic gauge symmetry are broken, but a combination
survives. This is similar to what happens in the standard electroweak
model, where both weak isospin and hypercharge are broken, but a
certain combination survives, to become electromagnetism as we know it. Just as
in that case, also in CFL+QED the charge spectrum is modified. One
finds that the quarks, gluons, and pseudoscalars acquire integral
charges (in units of the electron charge); in fact, the charges
match those of the corresponding hadrons precisely.

It is difficult to resist the conjecture that these two states are
continuously related to one another, with no phase transition, as the
density varies \cite{schaef1999}.  During this variation, degrees of 
freedom that are
``obviously" three-quark baryons evolve continuously into degrees of
freedom that are ``obviously" single quarks. This nifty trick is
possible because diquarks can be exchanged with the condensate freely.

If the core of a neutron star is described by the color-flavor locked
(CFL) phase, which seems plausible, it will be a transparent
insulator that partially reflects light -- like a diamond! This
particular consequence of the CFL phase is unlikely to be observed
any time soon, but we are working toward defining indirect signatures
in observable neutron star and supernova properties.

Unfortunately, existing numerical methods for calculating the
behavior of QCD converge very slowly at high density and low temperature. They
are totally impractical, even for the biggest and best modern
computers.    Developing usable algorithms for this kind of problem
is a most important open challenge.

\bigskip
\bigskip

\section{Consistency, Minimalism, Explanatory Method}

In the preceding sections I've illustrated how the foundational concepts of Yang-Mills theory 
are most clearly and directly exhibited in physical phenomena,
and how these foundations support applications that extend well beyond the starting concepts.  
There can be no doubt that Yang-Mills theory reveals leading principles of Nature's operating system.  
Now I'd like to reflect briefly on how it fits together with other leading principles, 
and what it indicates about the overall system.

\subsection{Gauge Symmetry and Consistency}

Is gauge symmetry an autonomous concept, logically independent of other leading principles of physics?  
On the contrary, it appears to be mandatory, in the theory of vector particles, 
to insure consistency with special relativity and quantum mechanics.  
For if the transverse parts of the vector field produce excitations that have a 
normal probabilistic interpretation (i.e., the square of their amplitude is the probability for their presence), 
then Lorentz invariance implies that the longitudinal parts produce excitations that are, 
in the jargon of quantum theory, ghosts.  That is to say, the square of their amplitudes is \emph{minus}
the probability for their presence, so that when we contemplate their production we are confronted with 
the specter of negative probabilities, which on the face of it are senseless.  
Gauge invariance saves the day by insuring that the longitudinal modes decouple, 
i.e. that transition amplitudes to excite such modes actually vanish.  
Thus gauge invariance is required, in order to insure that no physical process is assigned a negative probability.   

\subsection{Minimal Coupling, Predictivity, and More Consistency}

The remarkable predictive power of gauge symmetry arises only after it is combined with 
constraints of locality and minimal coupling.   It is by orchestrating these requirements that 
we are led to a specific, tight framework, which licenses only extremely few allowed parameters.

Why minimal coupling?    This requirement arises indirectly, as a consequence of 
the general difficulty of constructing consistent interacting relativistic quantum field theories in 
four space-time dimensions.  

The perturbative construction of such theories involves regularization and renormalization.  
If we want to have a closed theory, in which the limiting effect of the regulator 
can be absorbed into a finite number of renormalization parameters, we must restrict ourselves 
to renormalizable couplings, that is interaction vertices whose coefficients have non-negative 
mass dimension (in units with $\hbar = c = 1$, of course).   Otherwise the insertion of these vertices will, 
in general, be accompanied by positive powers of the cutoff mass, and taking that mass to infinity 
will be counterproductive.  This means that the vertices themselves must be constructed from fields 
whose total mass dimension does not exceed four.  This criterion, when combined with gauge invariance, 
greatly restricts the possibilities, and in particular enforces minimal coupling for the gauge fields.  

A slightly different perspective on renormalizability is associated with the philosophy of effective field theory.
According to this philosophy it is presumptuous, or at least unnecessarily committal, 
to demand that our theories be self-contained up to arbitrarily large energies.  
So we should not demand that the effect of a high-mass cutoff, which marks the breakdown of 
our effective theory, can be removed entirely.  Instead, we acknowledge that new degrees of freedom 
may open up at the large mass scale, and we postulate only that these degrees of freedom 
approximately decouple from low-scale physics.  By requiring that the effective theory 
they leave behind should be self-contained and approximately valid up to the high mass scale, 
we are then led to a similar ``effective'' veto, which outlaws 
\emph{quantitatively significant} nonrenormalizable couplings.   

Of course, this philosophy only puts off the question of consistency, 
passing that burden on to the higher mass-scale theory.   
Presumably this regress must end somewhere, either in a fully consistent quantum field theory 
or in something else (string theory?).  

ItÕs interesting, in any case, to explore the boundaries of consistency in quantum field theory.  
Perturbative renormalizability does not exhaust the issue.  
It is unlikely that perturbation theory converges in any interesting example, 
and it is not entirely clear how (or whether) we can use divergent series to extract well-defined, 
physically acceptable answers.  There is only one way known to regulate a quantum field theory 
without reference to perturbation theory, and that is by discretizing it, specifically by restricting 
the variables to a space-time lattice.  Having done this, then to arrive at continuum, 
Poincare invariant amplitudes we must remove the cutoff, employing a renormalization procedure 
to keep the content of the theory fixed as we do so.  This program only works out 
in a straightforward way for asymptotically free theories, where the effect of the mangled short-distance modes is
small and controllable.   

Summing it up, the only theories that are known to realize the basic concepts of special relativity 
and quantum mechanics fully are asymptotically free theories.  
Since asymptotic freedom can be checked in perturbation theory, one can survey 
the possibilities systematically.  Upon doing so, one finds that all non-trivial asymptotically free theories involve 
nonabelian gauge fields.  This provides, I think, an important part of the ultimate explanation 
for the central role of Yang-Mills theory 
in fundamental physics.

\subsection{Patterns of Explanation}

If there are to be simple explanations for complex phenomena, what form can they take?

One archetype is symmetry.   
In fundamental physics, especially in the twentieth century, symmetry has
been the most powerful and fruitful guiding principle. 
By tying together the description of physical behavior in many different circumstances  --  
at different places, at different times, viewed at different speeds and, of course, in different gauges!  -- 
it allows us to derive a wealth of consequences from our basic hypotheses.  
When combined with the principles of quantum theory, symmetry imposes 
very stringent consistency requirements, as we have discussed, leading to tight, predictive theories, 
of which Yang-Mills theory forms the archetype within the archetype.  

(In the present formulation of physics quantum theory itself appears as a set of independent principles, 
which loosely define a conceptual framework.  
It is not absurd to hope that in the future these principles will be formulated more strictly, 
in a way that involves symmetry deeply.)   

A different archetype, which pervades biology and cosmology, is the unfolding of a program.  
Nowadays we are all familiar with the idea that simple computer programs, 
unfolded deterministically according to primitive rules, can produce fantastically complicated patterns, 
such as the Mandelbrot set and other fractals; and with the idea that a surprisingly small library of DNA code 
directs biological development.   

These archetypes are not mutually exclusive.  Conway's Game of Life, for example, 
uses simple, symmetric, deterministic rules, always and everywhere the same; 
but it can, operating on simple input, produce extremely complex, yet highly structured output.   

In fundamental physics to date, we have mostly got along without having 
to invoke partial unfolding of earlier, primary simplicity as a separate explanatory principle.   In constructing a
working model of the physical world, to be sure, we require specification of initial conditions for the fundamental equations. 
But we have succeeded in paring these initial conditions down to a few parameters 
describing small departures from space-time homogeneity and
thermal equilibrium in the very early universe; and the roles of these two aspects of world-construction, 
equations and initial conditions, have remained pretty clearly separated.   
Whether symmetry will continue to expand its explanatory scope, giving rise to laws of such power 
that their solution is essentially unique, thus minimizing the role of initial conditions; 
or whether  ``fundamental'' parameters (e.g., quark and lepton masses and mixing angles) in fact depend
upon our position within an extended, inhomogeneous Multiverse, 
so that evolutionary and anthropic considerations will be unavoidable; or whether some deeper synthesis will somehow
remove the separation,  
is a great question for the future.  However these issues ultimately play out, 
it is surely not premature to celebrate the extensive contributions of symmetry in general, 
and of Yang-Mills theory in particular, in enriching our understanding of the physical world.

  \end{document}